\documentclass[12pt]{article}
\usepackage{amssymb,latexsym, amsmath}
\usepackage{graphicx}
\newcommand{\bfi}{\bfseries\itshape}
\newcommand{\singlespace}{\baselineskip 4.2333mm \parskip
4.2333mm}

\makeatletter
\@addtoreset{figure}{section}
\def\fps@figure{h, t}
\@addtoreset{table}{bsection}
\def\thetable{\thesection.\@arabic\c@table}
\def\fps@table{h, t}
\@addtoreset{equation}{section}

\makeatother

\begin{document}

\title{{\bf The Navier-Stokes-alpha model
of fluid turbulence}}
\author{
{\small Ciprian  Foias}
\\{\small Department of Mathematics}
\\{\small  Indiana University}
\\{\small  Bloomington, IN 47405, USA}
\\{\footnotesize email: foias@indiana.edu}
\and
{\small Darryl D. Holm}
\\{\small T-Division and CNLS, MS-B284}
\\{\small Los Alamos National Laboratory}
\\{\small Los Alamos, NM 87545, USA}
\\{\footnotesize email: dholm@lanl.gov}
\and
{\small Edriss S. Titi}
\\{\small Departments of Mathematics, Mechanical and Aerospace
Engineering}
\\{\small  University of California}
\\{\small Irvine, CA 92697, USA}
\\{\footnotesize email: etiti@math.uci.edu}
}
\date{
{\bfi Dedicated to V. E. Zakharov\\ 
on the occasion of his 60th birthday}\\
(To appear in {\it Physica D})}

\maketitle

%%%%%%%%%%%%%%%%%%%%%%%%%%%%%%%%%%%%%%%%%%%%%%%%%%%%%%%%%%%%%%%%

\begin{abstract}\noindent
We review the properties of the nonlinearly
dispersive Navier-Stokes-alpha (NS$-\alpha$) model
of incompressible fluid turbulence -- also called the viscous
Camassa-Holm equations in the literature. We first 
re-derive the NS$-\alpha$ model by filtering the velocity of
the fluid loop in Kelvin's circulation theorem for the
Navier-Stokes equations. Then we show that this filtering causes
the wavenumber spectrum of the translational kinetic energy for
the NS$-\alpha$ model to roll off  as $k^{-3}$ for
$k\alpha>1$ in three dimensions, instead of continuing
along the slower Kolmogorov scaling law, $k^{-5/3}$, that it 
follows for $k\alpha<1$. This rolloff at higher wavenumbers
shortens the inertial range for the NS$-\alpha$ model and thereby
makes it more computable. We also explain how the NS$-\alpha$
model is related to large eddy simulation (LES)
turbulence modeling and to the stress tensor for
second-grade fluids. We close by surveying recent results in the
literature for the NS$-\alpha$ model and its inviscid limit (the
Euler$-\alpha$ model). 
\end{abstract}

%%%%%%%%%%%%%%%%%%%%%%%%%%%%%%%%%%%%%%%%%%%%%%%%%%%%%%%%%%%%%%%%

\section{Introduction}
The energy in a turbulent fluid flow cascades toward ever
smaller scales until it reaches the dissipation scale, where it
can be transformed into heat. This cascade -- creating
fluid motions at ever smaller scales -- is a characteristic
feature of turbulence. This feature is also the main difficulty
in simulating turbulence numerically, because all numerical
simulations will have finite resolution and will eventually be
unable to keep up with the cascade all the way
to the dissipation scale, especially for complex
flows, e.g., near walls and interfaces.

The effects of subgrid-scale fluid motions occurring below the
available resolution of numerical simulations must be modeled.
One way of modeling these effects is simply to discard the energy
that reaches such subgrid scales. This is clearly unacceptable,
though, and many creative alternatives have been offered. A
prominant example is the large eddy simulation (LES) approach,
see, e.g., \cite{MGermano[1992]}, \cite{SandipGhosal[1999]}.
The LES approach is based on applying a spatial filter to the
Navier-Stokes equations. The reduction of flow complexity and
information content achieved in the LES approach depends on the
characteristics of the filter that one uses, its type and width.
In particular, the LES approach introduces a length scale into
the description of fluid dynamics, namely, the width of the
filter used. Note that the LES approach is conceptually
different from the Reynolds Averaged Navier-Stokes (or, RANS)
approach, which is based on statistical arguments and exact
ensemble averages, rather than spatial and temporal filtering.
After filtering, however, just as in the RANS approach, one faces
the classic turbulence closure problem: How to model the
filtered-out subgrid scales in terms of the remaining resolved
fields? In practice, this problem is compounded by the
requirement that the solution be simulated numerically, thereby
introducing further approximations.

This paper begins by reviewing a modeling scheme --- called here
the Navier-Stokes-alpha model, or NS$-\alpha$ model (also called
the viscous Camassa-Holm equations in \cite{Chen-etal-PRL[1998]} -
\cite{FHTi}$\,$) --- that introduces an energy
``penalty" inhibiting the creation of smaller and smaller
excitations below a certain length scale (denoted alpha). This
energy penalty results in a {\bfi nonlinearly dispersive
modification} of the  Navier-Stokes equations. The
alpha-modification appears in the nonlinearity, it depends on
length scale and we emphasize that it is dispersive, not
dissipative. We shall re-derive the modified equations in
Section 2 from the viewpoint of Kelvin's circulation theorem. As
we shall show in section 3, this modification causes the
translational kinetic energy wavenumber spectrum of the
NS$-\alpha$ model to roll off rapidly below the length scale
alpha as $k^{-3}$ in three dimensions, instead of continuing to
follow the slower Kolmogorov scaling law,
$k^{-5/3}$. This roll off shortens the inertial range of the
NS$-\alpha$ model and thus makes it more computable. This is the
main new result of the paper, so we shall comment now on 
its main implications.

Since the energy spectrum rolls off faster, the wavenumber
$k=\kappa_\alpha$ at which viscous dissipation takes over in the
NS$-\alpha$ model must be {\it lower} than for the original
Navier-Stokes equations. Hence, for a given driving force and
viscosity, the number of active degrees of freedom $N_{dof}$ for
the NS$-\alpha$ model must be {\it smaller} than for
Navier-Stokes. In section 3 we give a heuristic estimate of the
number $N_{dof}$ and compare it with the rigorous estimate
derived in \cite{FHTi} for the fractal dimension
$D_{frac}$ of the global attractor for the NS$-\alpha$ model.
Namely,
\begin{equation}
D_{frac} \leq (N_{dof})^{3/2}
\quad\hbox{and}\quad
N_{dof}
\equiv
 (L\kappa_\alpha)^3
\simeq
 \frac{L}{\alpha}Re^{3/2}
\,,
\end{equation}
where $L$ is the integral scale (or domain size), $\kappa_\alpha$
is the end of the NS$-\alpha$ inertial range and
$Re=L^{4/3}\epsilon_\alpha^{1/3}/\nu$ is the Reynolds number
(with NS$-\alpha$ energy dissipation rate $\epsilon_\alpha$ and
viscosity $\nu$).  The corresponding number of degrees of freedom
for a Navier-Stokes flow with the {\it same} parameters is
\begin{equation}
N_{dof}^{NS}
\equiv
 (L/\ell_{Ko})^3
\simeq
Re^{9/4}
\,,
\end{equation}
where $\ell_{Ko}$ denotes the Kolmogorov dissipation length scale.
The implication of these estimates of degrees of
freedom for numerical simulations that access a significant
number of them using the NS$-\alpha$ model would be an increase
in computational speed relative to Navier-Stokes of
\begin{equation}\label{factor-intro}
\bigg(\frac{N_{dof}^{NS}}{N_{dof}}\bigg)^{4/3}
=
 \bigg(\frac{\alpha}{L}\bigg)^{4/3}
\!\! Re\,.
\end{equation}
Thus, if $\alpha$ tends to a constant value, say $L/100$, when the
Reynolds number increases -- as found in
\cite{Chen-etal-PRL[1998]} - \cite{Chen-etal-PhysD[1998]} by
comparing steady NS$-\alpha$ solutions with experimental
data for turbulent flows in pipes and channels -- then one could
expect to obtain a substantial increase in computability by using
the NS$-\alpha$ model at high Reynolds numbers. An early
indication of the reliability of using these estimates to gain a
relative increase in computational speed in direct numerical
simulations (DNS) of homogeneous turbulence in a periodic domain
is given in \cite{Chen-etal[1999]}. There, a computational speed
up was achieved by using the NS$-\alpha$ model in DNS of
turbulence in a periodic domain by a factor about equal to
the fluctuation Reynolds number ($\approx250$). In the case
considered in \cite{Chen-etal[1999]}, this factor happens to be
about $Re/100$. 

In section 4, we discuss the relation of the NS$-\alpha$ model
to similar equations which are derived in a different physical
context, namely in the context of non-Newtonian fluids. The main
difference between the NS$-\alpha$ model and the non-Newtonian
second-grade fluids which it resembles lies in the choice of 
dissipation. The NS$-\alpha$ model uses the standard
Navier-Stokes viscosity, while the second-grade fluid uses a
weaker form of dissipation -- namely, wavenumber independent
damping of the fluid velocity. Discussions of the relative
advantages of the two forms of dissipation (e.g., in terms of
boundary data requirements and well-posedness) are beyond the
scope of this review.  However, we remark that the steady
solutions of the NS$-\alpha$ model with the standard
Navier-Stokes viscosity were found in \cite{Chen-etal-PRL[1998]}
- \cite{Chen-etal-PhysD[1998]} to agree with experimental mean
velocity profile data for turbulent flows in pipes and channels.
The corresponding steady solutions of the second-grade fluid with
its weaker form of dissipation were not found to so agree with
this experimental data. Finally, in section 5, we provide a brief
guide to the recent literature for those who might be interested
in the mathematical context in which the alpha model was
originally derived, as well as in its potential applications in
turbulence modeling. In regard to the latter, the NS$-\alpha$
model was recently shown to be transformable to a generalized
similarity model for large eddy simulation (LES) turbulence
modeling \cite{Dom-Holm[2000]}.

%%%%%%%%%%%%%%%%%%%%%%%%%%%%%%%%%%%%%%%%%%%%%%%%%%%%%%%%%%%%%%%%

\section{Kelvin-filtered turbulence models} Although it was
first derived from energetic considerations using the
Euler-Poincar\'e variational framework in \cite{HMR-98a},
\cite{HMR-98b}, the Navier-Stokes-alpha model may be motivated by
an equivalent argument based on Kelvin's circulation theorem. The
original Navier-Stokes (NS) equations are
\begin{equation}\label{NS-eqns}
\frac{\partial\mathbf{v}}{\partial t}
+ \mathbf{v}\cdot\nabla\mathbf{v} + \nabla p
= \nu\Delta\mathbf{v}
+ \mathbf{f}\,,
\quad\hbox{with}\quad
\nabla\cdot\mathbf{v}
= 0
\,,
\end{equation}
for a forcing $\mathbf{f}$ and constant kinematic viscosity
$\nu$. These equations satisfy {\bfi Kelvin's circulation
theorem},
\begin{equation}\label{KelThm-NS}
\frac{d}{dt}
\oint_{\gamma(\mathbf{v})}
\mathbf{v}\cdot d\mathbf{x}
=
\oint_{\gamma(\mathbf{v})}
(\nu\Delta\mathbf{v}
+ \mathbf{f})\cdot d\mathbf{x}
\,,
\end{equation}
for a fluid loop $\gamma(\mathbf{v})$ that moves with velocity
$\mathbf{v}(\mathbf{x},t)$, the Eulerian fluid velocity.

%%%%%%%%%%%%%%%%%%%%%%%%%%%%%%%%%%%%%%%%%%%%%%%%%%%%%%%%%%%%%%%%

\paragraph{Kelvin-filtering the Navier-Stokes equations.}
The equations for the NS$-\alpha$ model emerge from a
modification of the Kelvin circulation theorem (\ref{KelThm-NS})
to integrate around a loop $\gamma(\mathbf{u})$ that moves with a
{\bfi spatially filtered Eulerian fluid velocity} given
by $\mathbf{u}=g*\mathbf{v}$, where $*$ denotes the convolution,
\begin{equation}\label{g-star}
\mathbf{u}=g*\mathbf{v}
= \int g(\mathbf{x}-\mathbf{y})\,\mathbf{v}\
d^3y
\,.
\end{equation}
The ``inverse'' is denoted
\begin{equation}\label{g-inv}
\mathbf{v}=\mathcal{O}\mathbf{u}
\,,
\end{equation}
thereby defining an operator $\mathcal{O}$ whose Green's
function is the filter $g$ and which we shall assume is
positive, symmetric, isotropic and time-independent. Under
these assumptions the quantity (kinetic energy)
\begin{equation}\label{KE-norm}
E =
\frac{1}{2}\int
\mathbf{u}\cdot\mathbf{v}\ d^3x
=
\frac{1}{2}\int \mathbf{v}\cdot g*\mathbf{v}\
d^3x
= \frac{1}{2}\int\mathbf{u}\cdot\mathcal{O}\mathbf{u}\
d^3x
\,,
\end{equation}
defines a norm.

We obtain a modification to the Navier-Stokes equations
(\ref{NS-eqns}) by replacing in their Kelvin's circulation
theorem (\ref{KelThm-NS}) the loop $\gamma(\mathbf{v})$ with
another loop $\gamma(\mathbf{u})$ moving with the spatially
filtered velocity, $\mathbf{u}$. Then we have,
\begin{equation}\label{KelThm-NS-alpha}
\frac{d}{dt}
\oint_{\gamma(\mathbf{u})}
\mathbf{v}\cdot d\mathbf{x}
=
\oint_{\gamma(\mathbf{u})}
(\nu\Delta\mathbf{v}
+ \mathbf{f})\cdot d\mathbf{x}
\,.
\end{equation}
After taking the time derivative inside the Kelvin loop
integral moving with filtered velocity $\mathbf{u}$ and
reconstructing the gradient of pressure, we find the
{\bfi Kelvin-filtered Navier-Stokes equation},
\begin{equation}\label{NS-alpha-eqns}
\frac{\partial\mathbf{v}}{\partial t}
+ \mathbf{u}\cdot\nabla\mathbf{v}
+ \nabla\mathbf{u}^{\rm T}\cdot\mathbf{v}
+ \nabla p
= \nu\Delta\mathbf{v}
+ \mathbf{f}\,,
\end{equation}
with
\begin{equation}\label{aux-cond}
\nabla\cdot\mathbf{u}
= 0
\,,\quad\hbox{and}\quad
\mathbf{v}=\mathcal{O}\mathbf{u}
\,.
\end{equation}
The velocity $\mathbf{u}(\mathbf{x},t)$ is the spatially
filtered Eulerian fluid velocity in equation (\ref{g-star}).
Note that continuity equation is now imposed as
$\nabla\cdot\mathbf{u}= \nabla\cdot(g*\mathbf{v})=0$.
The energy balance relation derived from the Navier-Stokes-alpha
equations (\ref{NS-alpha-eqns}) is
\begin{equation}\label{KE-alpha-diss}
\frac{d}{dt}\int
\frac{1}{2}\,\mathbf{u}\cdot\mathbf{v}\ d^3x
=
\int
\mathbf{u}\cdot\mathbf{f}\ d^3x
-
\int
 \nu\, |\nabla\mathcal{O}^{1/2}\mathbf{u}|^2\ d^3x
\,,
\end{equation}
where for the moment we have dropped the boundary terms that
appear upon integrating by parts. That is,
for the moment we ignore the boundary effects and
consider either the case of the whole space with solutions
vanishing sufficiently rapidly at infinity, or the case of
periodic boundary conditions.

%%%%%%%%%%%%%%%%%%%%%%%%%%%%%%%%%%%%%%%%%%%%%%%%%%%%%%%%%%%%%%%%

\paragraph{Similarities with previous work.}
\begin{enumerate}
\item
Except for the term $(\nabla\mathbf{u})\cdot\mathbf{v}$, the
Kelvin-filtered Navier-Stokes equation (\ref{NS-alpha-eqns}) is
otherwise quite similar to Leray's regularization of the
Navier-Stokes equations proposed in 1934 \cite{Leray[1934]}.
Extension of the Leray regularization to satisfy the Kelvin
circulation theorem was cited as an outstanding problem in
Gallavotti's review \cite{Gallavotti[1992]}.
Looking at the equation of motion for the vorticity
$\mathbf{q} = \nabla \times \mathbf{v}$, reveals even more
similarity with Leray's regularization of the Navier-Stokes
equations. (See the section about the vorticity below.)
\item
At first glance, the Kelvin-filtered equations
(\ref{NS-alpha-eqns}) in the absence of dissipation and forcing
may seem reminiscent of a form of the Euler equations that was
discussed in Kuz'min \cite{Kuz'min-1983} and  Oseledets
\cite{Oseledets-1989} (see also Gama and Frisch
\cite{Gama-Frisch[1993]}), namely,
\begin{eqnarray}\label{KO-rep-Eul}
\frac{\partial\gamma_i}{\partial t}
+ u^j\frac{\partial\gamma_i}{\partial x^j}
&=&
-\, \gamma_j\frac{\partial u^j}{\partial x^i}
\,,
\nonumber\\
\gamma_i
&=& u_i + \frac{\partial\phi}{\partial x^i}\,,
\nonumber\\
\frac{\partial u_i}{\partial x^i}
&=&0
\,.
\end{eqnarray}
In light of the third equation in this set, the second one is
essentially the Hodge decomposition of the vector
$\gamma$ with Euclidean components
$\gamma_i$. Comparing these equations with Euler's equations
for the incompressible motion of an ideal fluid,
\begin{equation}\label{Eul-eqns}
\frac{\partial u_i}{\partial t}
+ u^j\frac{\partial u_i}{\partial x^j}
+\nabla p
= 0\,,
\quad
\frac{\partial u_i}{\partial x^i}
 = 0
\,,
\end{equation}
gives a relation between the pressure gradient and the ``gauge
function'' $\phi$ that is reminiscent of (but different from)
Bernoulli's law,
\begin{equation}\label{KO-pressure}
\nabla p
= \nabla\Big(\frac{1}{2}|\mathbf{u}|^2
- \frac{\partial\phi}{\partial t}
 - u^j\frac{\partial\phi}{\partial x^j}\Big)
\,.
\end{equation}
This type of relationship also arises in Hamilton's principle
when one uses Clebsch variables \cite{HK83}, in which case
$\phi$ is a Lagrange multiplier that enforces the continuity
equation. In contrast, equations (\ref{NS-alpha-eqns}) in the
ideal unforced case give the {\bfi Kelvin-filtered Euler model},
\begin{eqnarray}\label{Euler-Kel-model}
\frac{\partial v_i}{\partial t}
+ u^j\frac{\partial v_i}{\partial x^j}
&=&
-\, v_j\frac{\partial u^j}{\partial x^i}
- \frac{\partial p}{\partial x^i}
\,,
\\
v_i
&=& \mathcal{O}u_i
\,,
\nonumber\\
\frac{\partial u_i}{\partial x^i}
&=&0
\,.
\nonumber
\end{eqnarray}
The Kuz'min-Oseledets form of the Kelvin-filtered Euler model
appears by replacing $u_i\to v_i$ only in the second of the
three equations in (\ref{KO-rep-Eul}).

\end{enumerate}

%%%%%%%%%%%%%%%%%%%%%%%%%%%%%%%%%%%%%%%%%%%%%%%%%%%%%%%%%%%%%%%%

\paragraph{Conservation laws.}
References to the interesting geometrical
properties of the Euler$-\alpha$ model equations -- namely
equations (\ref{Euler-Kel-model}) when $\mathcal{O}$ is the
Helmholtz operator -- are cited in the last Section. At
this point, we only comment that in the case of periodic boundary
conditions, or the case of the whole space with solutions
vanishing sufficiently rapidly at infinity, these nondissipative
equations preserve the following two quadratic invariants, the
{\bfi kinetic energy},
\begin{equation}\label{KE-Eul-alpha}
E =
\frac{1}{2}\int
\mathbf{u}\cdot\mathbf{v}\,d^3x
\,,
\end{equation}
and the $\mathbf{v}-$ {\bfi helicity},
\begin{equation}\label{Eul-alpha-helicity}
\Lambda = \int
\mathbf{v}\cdot\hbox{curl}\,\mathbf{v}\,d^3x.
\end{equation}
The kinetic energy, $E_\alpha$, and the $\mathbf{v}$-helicity,
$\Lambda$, defined above, are also preserved in bounded domains
 provided appropriate boundary conditions are imposed.
Boundary conditions sufficient for energy conservation when
$\mathcal{O}$ is the Helmholtz operator $1-\alpha^2\Delta$ are
\begin{equation}\label{alpha-bc}
\hat{\mathbf{n}}\times\Big(\hat{\mathbf{n}}\cdot
(\nabla\mathbf{u}+\nabla\mathbf{u}^T\,) \Big)
= 0
\,,
\end{equation}
where the vector $\hat{\mathbf{n}}$ is normal to
the boundary and superscript $(\cdot)^T$ denotes matrix transpose.
Boundary conditions sufficient for $\mathbf{v}$-helicity
conservation in general are
$\lambda\,\hat{\mathbf{n}}\cdot\mathbf{u}=0$ and
$p\,\hat{\mathbf{n}}\cdot\hbox{curl}\,\mathbf{v}=0$, as seen from
the helicity equation,
\begin{equation}\label{Eul-alpha-helicity-dyn}
\frac{\partial \lambda}{\partial t}
=
-\,\hbox{div}
\,(\lambda\,\mathbf{u}+p\,\hbox{curl}\,\mathbf{v})
\,,\quad\hbox{where}\quad
\lambda\equiv\mathbf{v}\cdot\hbox{curl}\,\mathbf{v}
\,.
\end{equation}
This equation is obtained by using {\it only} the motion equation
in  (\ref{Euler-Kel-model}) and its curl. Therefore,
conservation of helicity holds with these boundary conditions
simply because of the {\it form} of the Kelvin-filtered motion
equation in (\ref{Euler-Kel-model}), independently of the
relation between $\mathbf{v}$ and $\mathbf{u}$, and regardless of
whether $\mathbf{u}$ is incompressible.

%%%%%%%%%%%%%%%%%%%%%%%%%%%%%%%%%%%%%%%%%%%%%%%%%%%%%%%%%%%%%%%%

\paragraph{Vortex transport and stretching.} Let
$\mathbf{q}=\nabla\times\mathbf{v}$ be the vorticity of the
unfiltered Eulerian velocity. The curl of the Kelvin-filtered
Navier-Stokes equation (\ref{NS-alpha-eqns}) gives the vortex
transport and stretching equation,
\begin{equation}\label{NS-alpha-vortex}
\frac{\partial\mathbf{q}}{\partial t}
+ \mathbf{u}\cdot\nabla\mathbf{q}
- \mathbf{q}\cdot\nabla\mathbf{u}
= \nu\Delta\mathbf{q}
+ \nabla\times\mathbf{f}\,.
\end{equation}
We note that the coefficient $\nabla\mathbf{u}$ in the vortex
stretching term is the gradient of the {\bfi spatially filtered}
Eulerian velocity $\mathbf{u}$. Thus, Kelvin-filtering tempers
the vortex stretching in the modified Navier-Stokes equations
(\ref{NS-alpha-eqns}), while preserving the original form of the
vortex dynamics. This tempered vorticity stretching is also 
reminiscent of Leray's approach \cite{Leray[1934]} for
regularizing the Navier-Stokes equations.

Indeed, when $\mathcal{O}$ is the Helmholtz  operator we take 
full advantage of this regularized vorticity stretching effect to
prove in \cite{FHTi} the global existence and uniqueness of
strong solutions for the three dimensional NS$-\alpha$ model. In
fact, when the forcing is in a certain Gevrey class of
regularity (real analytic), then the NS$-\alpha$
solutions are also Gevrey regular, so that the Fourier
coefficients of the solution decay exponentially fast,
with respect to the wavenumbers, at a rate
that increases as $\alpha^2$ increases. See
\cite{Foias-Temam[1989]}, \cite{Doering-Titi[1995]} for the
corresponding analysis of the Gevrey properties of the
Navier-Stokes equations. This result suggests that the filtering,
or smoothing of the velocity $\mathbf{u}$ due to the presence of
alpha in the NS$-\alpha$ model could enhance the decay of its
wavenumber spectrum and produce an earlier observation of this
exponentially falling tail when the forcing is sufficiently
smooth, even when the viscosity is small.

%%%%%%%%%%%%%%%%%%%%%%%%%%%%%%%%%%%%%%%%%%%%%%%%%%%%%%%%%%%%%%%%

\paragraph{Specializing to the Navier-Stokes-alpha model.} The
special case of the NS$-\alpha$ model emerges from the
Kelvin-filtered Navier-Stokes equation (\ref{NS-alpha-eqns}) when
we choose the operator $\mathcal{O}$ to be the Helmholtz
operator, thereby introducing a constant $\alpha$ that has
dimensions of length,
\begin{equation}\label{alpha-op}
\mathcal{O}
=
1-\alpha^2\Delta\,,
\quad\hbox{with}\quad
\alpha=const
\,.
\end{equation}
In this case, the filtered and unfiltered fluid velocities in
equation (\ref{g-inv}) are related by
\begin{equation}\label{g-inv-alpha}
\mathbf{v}=(1-\alpha^2\Delta)\mathbf{u}
\,.
\end{equation}
The {\bfi Navier-Stokes-alpha model} is given by the
Kelvin-filtered Navier-Stokes equation (\ref{NS-alpha-eqns}) with
definition (\ref{g-inv-alpha}). The original derivation of the
ideal {\bfi Euler-alpha model} (the $\nu=0$ case of the
NS$-\alpha$ model) obtained by using the Euler-Poincar\'e
approach is given in {\cite{HMR-98a}, \cite{HMR-98b}. The
physical interpretations of
$\mathbf{u}$ and $\mathbf{v}$ as the Eulerian and Lagrangian mean
velocities are given in \cite{DDH-fluct[1999]}.

The corresponding kinetic energy norm (\ref{KE-norm}) for the
NS$-\alpha$ model is given by
\begin{equation}\label{KE-alpha}
E_\alpha = \int  \bigg[
\frac{1}{2}|\mathbf{u}|^2
+ \frac{\alpha^2}{2}|\nabla\mathbf{u}|^2 \bigg] d^3x
\,,
\end{equation}
and we repeat that $\mathbf{u}$ is the spatially filtered
Eulerian fluid velocity. This kinetic energy is the sum of a
translational kinetic energy based on the filtered velocity
$\mathbf{u}$, and a gradient-velocity kinetic energy, multiplied
by $\alpha^2$. Thus, by showing the global boundedness, in time,
of the kinetic energy  (\ref{KE-alpha}) one concludes that the
coefficient $\nabla\mathbf{u}$ in the vortex stretching relation
(\ref{NS-alpha-vortex}) is {\it bounded in} $L^2$ for the
NS$-\alpha$ model. The second term in the  kinetic energy
(\ref{KE-alpha}) imposes an energy penalty for creating small
scales. The spatial integral of $|\nabla\mathbf{u}|^2$ in the
second term has the same dimensions as filtered enstrophy (the
spatial integral of $| \nabla \times \mathbf{u}|^2$, the squared
filtered vorticity).
For a domain with boundary the spatial integral of
$|\nabla\mathbf{u}|^2$ in the second term in the kinetic energy
norm (\ref{KE-alpha}) is replaced by the integral of
${\rm trace}\,(4{\bf D}\cdot{\bf D})$, where ${\bf D}$ is
the strain rate tensor, ${\bf D}=(1/2)(\nabla {\bf u}+\nabla {\bf
u}^{\rm T})$ in Euclidean coordinates
\cite{Shkoller-preprint}. For the case we treat here, in
Euclidean coordinates and in the absence of boundaries, all of
these norms are equivalent for an incompressible flow.

%%%%%%%%%%%%%%%%%%%%%%%%%%%%%%%%%%%%%%%%%%%%%%%%%%%%%%%%%%%%%%%%

\section{Spectral scaling for the NS$-\alpha$ model}

\paragraph{A preliminary scaling argument.}
Scaling ideas originally due to Kraichnan
\cite{Kraichnan[1967]} for the case of two-dimensional
turbulence may be applied to estimate the effect of the second
term in the NS$-\alpha$ model energy (\ref{KE-alpha}) on the
energy spectrum in the present case. For sufficiently small
wavenumbers ($k\alpha\ll1$, but $kL\gg1$) the first term in the
energy $E_\alpha$ in (\ref{KE-alpha}) dominates and the second
term may be neglected. In this wavenumber region, the standard
Kolmogorov scaling argument for turbulence gives a
$k^{-5/3}$ spectrum by the usual dimensional argument for the
inertial range,
\begin{equation}\label{Kolmogorov-scaling}
\!\!\!\!
k = [L]^{-1}
\,,\quad
\epsilon_\alpha = [L]^2 [T]^{-3}
\,,\quad
E_\alpha(k) = [L]^3 [T]^{-2} = \epsilon_\alpha^a k^b
\,,
\ \hbox{if}\quad k\alpha\ll1
\,,
\end{equation}
so that $a=2/3$ and $b=-5/3$. Conversely, for
sufficiently large wavenumbers ($k\alpha\gg1$) and the second term
dominates in $E_\alpha$.  A preliminary scaling argument would
then indicate a $k^{-3}$ spectrum for $E_\alpha$ in the large
wavenumber region, according to
\begin{equation}\label{Kraichnan-scaling}
k = [L]^{-1}
\,,\quad
\eta_\alpha = [T]^{-3}
\,,\quad
E_\alpha(k) = [L]^3 [T]^{-2} = \eta_\alpha^a k^b
\,,
\ \hbox{if}\quad k\alpha\gg1
\,,
\end{equation}
so that $a=2/3$ and $b=-3$ in this case. Here $\eta_\alpha$ is the
rate of dissipation of the $\int |\nabla \mathbf u|^2$ part of the
kinetic energy (with the same dimensions as enstrophy).
Therefore, in the wavenumber region near $k\alpha\sim1$ the
kinetic energy spectrum may be expected to have a break in slope
and roll off from $k^{-5/3}$ to $k^{-3}$ scaling
\cite{Eyink-Thomson}.

The expectation from this preliminary scaling argument seems to
be confirmed in direct numerical simulations of the NS$-\alpha$
model, as shown in Figure 1 below taken from
\cite{Chen-etal[1999]}. Figure 1 shows the energy spectra
resulting from three DNS of the NS$-\alpha$ model with mesh sizes
of $256^3$ for three cases: with $\alpha/L = 0$ (the Navier-Stokes
equations), $1/32$ and $1/8$ for the same viscosity $\nu = 0.001$.
The corresponding Taylor microscale Reynolds numbers
$R_{\lambda}$ are reported as $147, 182$ and $279$,
respectively. The higher $\alpha$ (higher Reynolds number)
flows are found numerically to have {\bfi more compact energy
spectra}.
\begin{figure}
\begin{center}
\includegraphics[scale=0.7,angle=0]{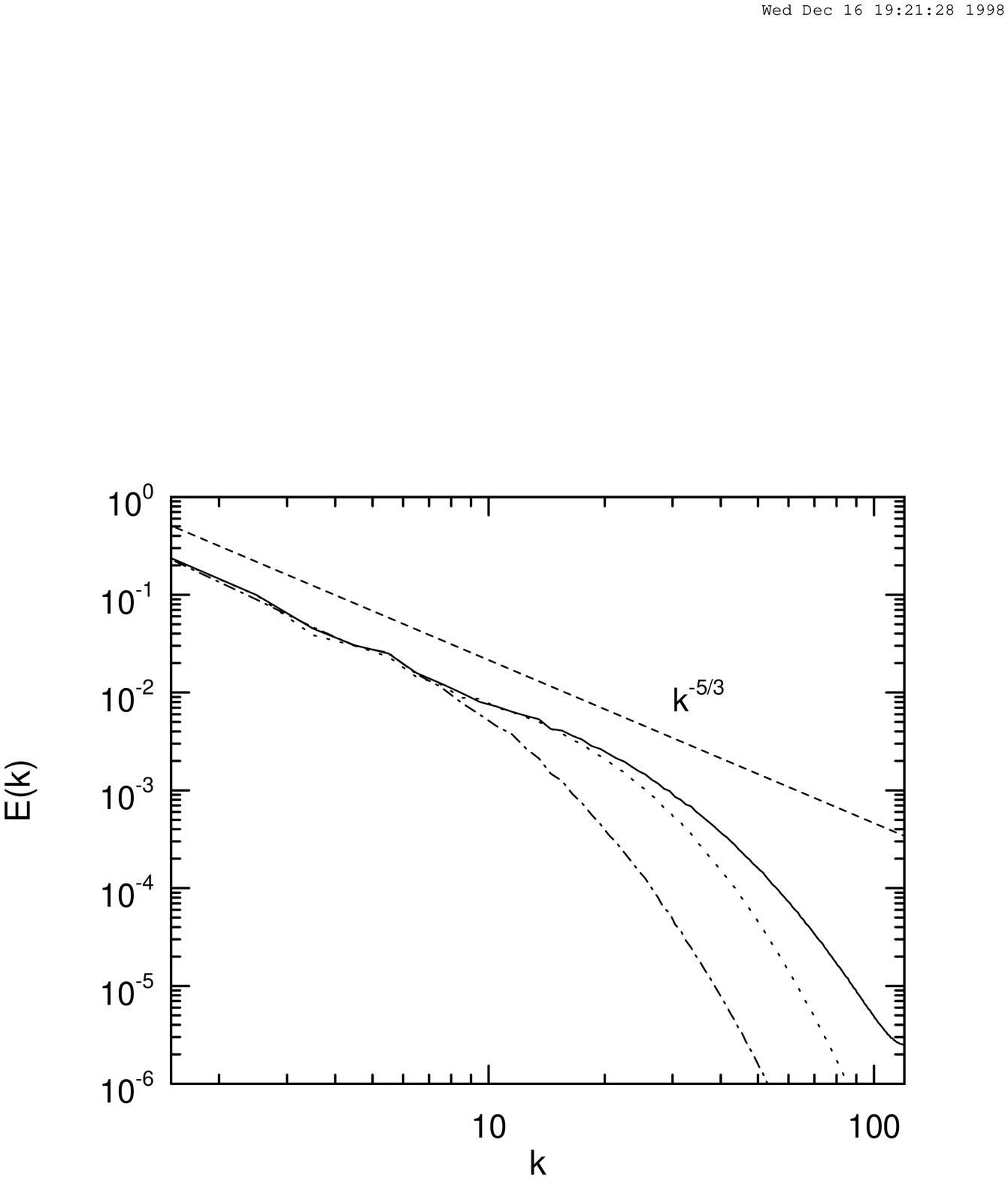}
\end{center}
\caption{\label{cp}{\footnotesize
The DNS energy spectrum, $E(k)=E_{\alpha}(k)$, versus the wave number $k$ for
three cases with the same viscosities, same forcings and mesh
sizes of $256^3$ for $\alpha = 0$ (solid line), $1/32$ (dotted
line) and $1/8$ (dotted-dash line). In the inertial range ($k <
20$), a power spectrum with $k^{-5/3}$ can be identified. For
finite $\alpha$, this behavior is seen to roll off to a steeper
spectrum for $k\ge1/\alpha$.}}
\end{figure}
%

%%%%%%%%%%%%%%%%%%%%%%%%%%%%%%%%%%%%%%%%%%%%%%%%%%%%%%%%%%%%
\paragraph{A more refined argument.}
A more refined argument for the  spectral scaling of
the NS$-\alpha$ model shall now be given that depends explicitly
upon properties of the nonlinearity.
Following \cite{Foias97}, let $u_\kappa$ denote the component of
$\mathbf{u}$ formed by the Fourier modes of fluid velocity
$\mathbf{u}$ with wavenumbers in $[\kappa,2\kappa)$ and similarly
let $v_\kappa$ denote the corresponding component of
$\mathbf{v}$, where $\kappa $ is well beyond the active
wavenumbers of the driving force.  The NS$-\alpha$ energy balance
for this component is then
\begin{equation}
\frac{1}{ 2} \frac{d}{dt} (u_\kappa ,v_\kappa )
+
 \nu (-\Delta
u_\kappa ,v_\kappa ) = T_\kappa -T_{2\kappa }
\,,
\label{(1)}
\end{equation}
where $T_\kappa $ and $T_{2\kappa }$ denote, respectively, the energy
transfer rate at $\kappa$ from the low wavenumber modes $u_<$ to the high
wavenumber modes $u_\kappa +u_>$, and at $2\kappa$ from $u_< +
u_{2\kappa}$ to $u_>$, where $u_<$ and $u_>$ are defined as
\begin{equation}
u_< \equiv \sum_{j<\kappa}u_j
\,,\quad\hbox{and}\quad
u_> \equiv \sum_{j\ge2\kappa}u_j
\,.
\end{equation}
In particular,
\begin{equation}
T_\kappa = -(\tilde B(u_<,v_<),u_\kappa ) + (\tilde
B(u_\kappa +u_>,v_\kappa +v_>),u_<)
\,,
\end{equation}
where the bilinear operator $\tilde B(u,v)$ is given by
\begin{equation}
\tilde B(u,v) = -P_\sigma \Big(u\times(\nabla\times v)\Big)
\,,
\end{equation}
and $P_\sigma$ is the $L^2$ orthogonal projection (Leray
projection), see, e.g., \cite{FHTi} for more mathematical
details.  For equation (\ref{(1)}) to hold, it is essential that
$\tilde B(u,v)$ has the property,
\begin{equation}
(\tilde B(u,v),u) = 0
\,.
\label{B-tilde-orthog}
\end{equation}
Note: property (\ref{B-tilde-orthog}) does not hold for the {\it
enstrophy} transfer rate of the NS$-\alpha$ model, so the
NS$-\alpha$ model does not possess a Kraichnan type inertial
range due to an enstrophy cascade. Time-averaging equation
(\ref{(1)}) gives
\begin{equation}
\nu \Big\langle(-\Delta u_\kappa ,v_\kappa )\Big\rangle
=
\langle\,T_\kappa\,\rangle - \langle\, T_{2\kappa}\,\rangle
.
\label{avgd-erg-rate}
\end{equation}
Consequently, upon introducing the energy spectrum
$E_\alpha(\kappa)\,,$ this time-averaged energy transfer
equation implies
\begin{equation}
\nu \kappa ^3E_\alpha (\kappa) \sim \nu \int^{2\kappa }_\kappa
\kappa ^2 E_\alpha (\kappa )d\kappa
\sim
\langle\,T_\kappa\,\rangle - \langle\, T_{2\kappa}\,\rangle.
\end{equation}
Hence, as long as the energy dissipation rate is small compared to the
energy transfer rate, i.e., provided
\begin{equation}
\nu \kappa ^3 E_\alpha (\kappa ) \ll \langle T_\kappa \rangle
,
\label{(2)}
\end{equation}
we have the inertial range condition,
\begin{equation}
\langle T_\kappa \rangle \sim \langle T_{2\kappa }\rangle
.
\end{equation}
The total energy dissipation rate is estimated as
\begin{equation}
\epsilon_\alpha  = \left\langle\frac{\nu}{L^3}
\int_{[0,L]^3}  u \cdot (-\Delta{v})
\,d^3x\right\rangle
.
\end{equation}
Kraichnan \cite{Kraichnan[1967]} posits the following mechanism
for the turbulent cascade: In the inertial range
the eddies
$u_\kappa $ transfer their energy to the eddies $u_{2\kappa }$ in
the time $\tau_\kappa$ it takes to travel their length $\sim
1/\kappa$.  Their average velocity being
\begin{equation}\label{U-kappa}
U_\kappa
\equiv
\bigg\langle \frac{1}{L^3}
\int_{[0,L]^3} \!\!\!\!
u_\kappa \cdot u_\kappa\,  d^3 x
\bigg\rangle^{1/2}\!\!\!\!\!\!
=
\bigg(
\int_\kappa^{2\kappa}\!\!
\frac{E_\alpha (\kappa)d\kappa}{(1+\alpha^2\kappa^2)}
\bigg)^{1/2}\!\!\!\!\!\!
\sim\
\bigg(
\frac{ \kappa E_\alpha (\kappa)}{(1+\alpha^2\kappa^2)}
\bigg)^{1/2}
\!\!\!\!\!\!,
\end{equation}
the eddy energy exchange, or turnover time $\tau_\kappa $ will
be
\begin{equation}\label{turnover}
\tau_\kappa  \sim 1/(\kappa U_\kappa)
\,.
\end{equation}
Consequently, the total energy dissipation rate is related to the
$\kappa$ spectral energy density by using (\ref{U-kappa}) as
\begin{equation}\label{Erg-dissip-rate}
\epsilon_\alpha
\sim
\tau_\kappa^{-1} \int^{2\kappa }_\kappa E_\alpha(k) dk
\sim
\kappa U_\kappa  \int^{2\kappa }_\kappa E_\alpha(k) dk
\sim
\frac{\kappa ^{5/2}}{(1+\alpha^2\kappa^2)^{1/2}}
 E_\alpha (\kappa )^{3/2}
\,,
\end{equation}
which yields the following spectral scaling law for the
NS$-\alpha$ inertial range,
\begin{equation}
E_\alpha (\kappa )
 \sim
\frac{\epsilon^{2/3}_\alpha (1+\alpha^2\kappa^2)^{1/3}}
{\kappa ^{5/3}}\,.
\label{(3)}
\end{equation}
Thus, the total energy present in the NS$-\alpha$ inertial range
is actually enhanced by the presence of alpha.  Notice, however,
that in terms of the filtered velocity $\mathbf{u}$ alone, the
{\bfi translational kinetic energy spectrum} is given by
\begin{equation}\label{E-alpha-k}
\frac{E_\alpha (\kappa )}{ (1+\alpha^2\kappa ^2)}
\sim
\frac{\epsilon_\alpha^{2/3}}{\kappa ^{5/3}}
\frac{1}{(1+\alpha^2\kappa ^2)^{2/3}}
=
\left\{
\begin{array}{ll}
{\displaystyle
\frac{ \epsilon_\alpha^{2/3} }{ \kappa^{5/3} }
}
& \hbox{if}\quad \kappa\alpha\ll1
\,,
\\
\\
{\displaystyle
\frac{\epsilon_\alpha^{2/3} }{ \alpha^{4/3}\kappa^3 }
}
&\hbox{if}\quad \kappa\alpha \gg1
\,.
\end{array}
\right.
\end{equation}
Hence, the anticipated $\kappa^{-3}$ behavior appears in the
spectrum for the $\alpha-$filtered velocity as a consequence of
equation (\ref{E-alpha-k}) arising from Kraichnan's argument
\cite{Kraichnan[1967]} associating energy transfer rates and
eddy turnover times for the filtered velocity. According
to this argument, the presence of alpha reduces the
energy associated with the higher wave numbers in the
$L^2$ norm of the filtered velocity $\mathbf{u}$. This
conclusion from (\ref{E-alpha-k}) agrees with the trends
shown in Figure 1 obtained from high resolution DNS studies of
the NS$-\alpha$ equation in \cite{Chen-etal[1999]}. However,
these numerical studies do not have sufficient dynamic range to
confirm this conclusion entirely. Thus, uncertainty remains in
claiming a numerical confirmation because the quantity
$E_\alpha(\kappa)/(1+\alpha^2\kappa ^2)$ in equation
(\ref{E-alpha-k}) is unaffected by the alpha-modification for
$\kappa\alpha\ll1$ and may already be out of the inertial
range and into the dissipation range for $\kappa\alpha\gg1$.
Studies in progress using DNS of the limit $\alpha\to\infty$
NS$-\alpha$ equation hope to clarify this point
\cite{Chen-etal-00}. Next we shall discuss the extent of the
inertial range for the NS$-\alpha$ model.

Equation ({\ref{E-alpha-k}) holds only
in the inertial range, that is, provided, cf. (\ref{(2)}),
\begin{equation}
\nu \kappa ^3 E_\alpha (\kappa ) \ll \epsilon_\alpha
\,.
\end{equation}
This inertial-range inequality may be expressed equivalently as
$\kappa\ll\kappa_\alpha$ where $\kappa _\alpha$ (the end of the
NS$-\alpha$ inertial range) is given in terms of the NS$-\alpha$
Kolmogorov wavenumber $\kappa_{\alpha ,Ko}$ by using
(\ref{Erg-dissip-rate}) and (\ref{E-alpha-k}) to find
\begin{equation}\label{k-alpha-rel}
\kappa_\alpha ^4 (1+\alpha ^2\kappa ^2_\alpha )
\sim
\frac{\epsilon_\alpha }{ \nu ^3}
=
 \kappa ^4_{\alpha ,Ko}
\,.
\end{equation}
Thus, as expected from the numerical simulations of
\cite{Chen-etal[1999]}, {\bfi the inertial range is shortened} to
$\kappa<\kappa_\alpha$  for the NS$-\alpha$ model by its
nonlinear dispersive filtering with lengthscale $\alpha$.  For
sufficiently large NS$-\alpha$ Kolmogorov wavenumber
$\kappa_{\alpha ,Ko}$ and with
$\alpha$ fixed, the  wavenumber $\kappa_\alpha$ at the end of
the NS$-\alpha$ inertial range is determined from
(\ref{k-alpha-rel}) to be
\begin{equation}
\kappa _\alpha
\sim
\left(\frac{1}{\alpha}\right)^{1/3}
\kappa ^{2/3}_{\alpha ,Ko}
\,.
\end{equation}
This is a relationship among the three progressively larger
wavenumbers,
$$
1/\alpha < \kappa_\alpha < \kappa_{\alpha,Ko}.
$$
Shortening the inertial range for the NS$-\alpha$ model to
$\kappa<\kappa_\alpha$ rather than $\kappa<\kappa_{\alpha,Ko}$
implies fewer active degrees of freedom in the solution.

%%%%%%%%%%%%%%%%%%%%%%%%%%%%%%%%%%%%%%%%%%%%%%%%%%%%%%%%%%%%%%

\paragraph{Counting degrees of freedom.}
If one expects turbulence to be ``extensive'' in the
thermodynamic sense, then one may expect that the
number of ``active degrees of freedom'' $N_{dof}$ for alpha-model
turbulence should scale as
\begin{equation}\label{NS-alpha-dof}
N_{dof}
\sim (L\kappa _\alpha )^3
\sim (L/\alpha ) (L\kappa_{\alpha,Ko})^2
\sim
 \frac{L}{\alpha}Re^{3/2}
\,,
\end{equation}
where $L$ is the integral scale (or domain size), $\kappa_\alpha$
is the end of the NS$-\alpha$ inertial range and
$Re=L^{4/3}\epsilon_\alpha^{1/3}/\nu$ is the Reynolds number
(with dissipation rate $\epsilon_\alpha$ and viscosity $\nu$).
The corresponding number of degrees of freedom for
Navier-Stokes with the {\it same} parameters is
\begin{equation}\label{NS-dof}
N_{dof}^{NS}
\equiv
 (L\,\kappa_{\alpha,Ko})^3
\sim
Re^{9/4}
\,,
\end{equation}
and one sees a possible trade-off in the relative Reynolds number
scaling of the two models. Should these estimates of the number of
degrees of freedom needed for numerical simulations using the
NS$-\alpha$ model relative to Navier-Stokes not be overly
optimistic, the implication would be one factor of
$({N_{dof}^{NS}/N_{dof}})^{1/3}$ in relative increased
computational speed gained by the NS$-\alpha$ model for each
spatial dimension and yet another factor for the
accompanying lessened CFL time step restriction. Altogether,
this would be a gain in speed of
\begin{equation}\label{factor}
\bigg(\frac{N_{dof}^{NS}}{N_{dof}}\bigg)^{4/3}
=
 \bigg(\frac{\alpha}{L}\bigg)^{4/3}
\!\! Re\,.
\end{equation}
Since $\alpha/L\ll1$ and $Re\gg1$, the two factors in the last
expression compete, but the Reynolds number should eventually win
out, because $Re$ can keep increasing while the number $\alpha/L$
is expected to tend to a constant value, say $\alpha/L=1/100$, at
high (but experimentally attainable) Reynolds numbers, at least
for simple flow geometries. Empirical indications for this
tendency were found in \cite{Chen-etal-PRL[1998]} -
\cite{Chen-etal-PhysD[1998]} by comparing steady NS$-\alpha$
solutions with experimental mean-velocity-profile data for
turbulent flows in pipes and channels.

Thus, according to this scaling argument, a factor of $10^4$ in
increased computational speed for resolved scales greater than
$\alpha$ could occur by using the NS$-\alpha$ model at the
Reynolds number for which the ratio
$\kappa_{\alpha,Ko}/\kappa_\alpha=10$. An early indication of
the feasibility of obtaining such factors in increased
computational speed was realized in the direct numerical
simulations of homogeneous turbulence reported in
\cite{Chen-etal[1999]}, in which
$\kappa_{\alpha,Ko}/\kappa_\alpha\simeq4$ and the full factor of
$4^4=256$ in computational speed was obtained using spectral
methods in a periodic domain at little or no cost of accuracy in
the statistics of the resolved scales.

%%%%%%%%%%%%%%%%%%%%%%%%%%%%%%%%%%%%%%%%%%%%%%%%%%%%%%%%%%%%%%%%

\paragraph{Related mathematical results.} The paper \cite{FHTi}
shows that strong solutions of the NS$-\alpha$ model exist
globally, they are unique, and they lie on a global attractor
whose fractal (Lyapunov) dimension is bounded above by
\begin{equation}\label{FractalDimBd}
\mathcal{D}_{frac} \le N_{dof}^{3/2}
\,,
\end{equation}
with $N_{dof}$ defined as in equation (\ref{NS-alpha-dof}).
This rigorous mathematical bound exceeds the expected value
obtained above by heuristically counting degrees of freedom
assuming ``extensive'' turbulence. Thus, it would imply a
smaller increase in computational speed than that of
(\ref{factor}). However, this rigorous bound may have room for
improvement.

%%%%%%%%%%%%%%%%%%%%%%%%%%%%%%%%%%%%%%%%%%%%%%%%%%%%%%%%%%%%%%%%

\paragraph{Oboukov cascade rate.}
We have established in equation~(\ref{turnover})  that the eddy
turnover rate based on the translational velocity for the
NS$-\alpha$ model is 
\begin{equation}
\tau_\kappa^{-1}
\sim
\kappa U_\kappa 
\sim 
\sqrt{\frac{\kappa^3 E_{\alpha}(k)}{1+\alpha^2 \kappa^2}} 
\sim
 \left(\frac{\epsilon_\alpha \kappa^2}
{1+\alpha^2 \kappa^2}\right)^{1/3}
\!\!.
\end{equation}
Therefore, 
\begin{itemize}
\item for low wavenumbers ($\kappa\alpha \ll 1$), information (or
error) propagates between scales $\kappa$ and $2\kappa$ at a rate
proportional to $\kappa^{2/3}$, in agreement with the classical
{\bfi accelerated cascade} of Oboukov \cite{Oboukov} as cited,
e.g., in \cite{Lesieur}; while 
\item for high wavenumbers ($\kappa\alpha \gg 1$), information
propagates at a constant rate, {\bfi independently of wavenumber}
for the NS$-\alpha$ model  (as occurs also in the case of 2D
turbulence discussed by Leith and Kraichnan
\cite{Leith-Kraichnan-70's}).
\end{itemize}
Without an accelerated cascade at high wavenumber, the
NS$-\alpha$ model should tend to be {\bfi more predictable} than
the original Navier-Stokes model. Of course, it makes sense that
an averaged, or filtered description of  fluid flow would
propagate high wavenumber information and errors at a slower rate
than an instantaneous, unfiltered description does (e.g., think of
climate vs weather).  The $\kappa^{-3}$ spectrum of the
translational kinetic energy in the alpha models for
$\alpha\kappa \gg 1$ is consistent with this interpretation of
{\bfi reduced error propagation rate} as measured in the $L^2$
norm of the filtered velocity.

%%%%%%%%%%%%%%%%%%%%%%%%%%%%%%%%%%%%%%%%%%%%%%%%%%%%%%%%%%%%%%%%
\section{Rheology of NS$-\alpha$ turbulence: 2nd-grade
fluids} We rewrite the NS-alpha equations (\ref{NS-alpha-eqns})
with
${\bf v} = (1 - \alpha^2\Delta){\bf u}$ and $\alpha$ constant in
their equivalent constitutive form,
\begin{equation}
\label{VCHE-tnsr}\hspace{-.35in}
	\frac{d{\bf u}}{dt}=\nabla\cdot {\bf T}\,,
 \quad\hbox{where}\quad
	{\bf T}=-p{\bf I}+2\nu(1 - \alpha^2\Delta){\bf D}
	+2\alpha^2 {\bf D}^{^{^{\hspace{-3mm}\small{\circ}}}}\
,
\end{equation}
with $\nabla\cdot {\bf u}=0$, strain rate ${\bf D}=(1/2)(\nabla
{\bf u}+\nabla {\bf u}^{\rm T})$, vorticity tensor
${\Omega}=(1/2)(\nabla{\bf u}-\nabla{\bf u}^{\rm T})$,
and co-rotational (Jaumann) derivative given by
${\bf D}^{^{^{\hspace{-3mm}\small{\circ}}}} = {d}{\bf
D}/{dt}+{\bf D}{\Omega}-{\Omega}{\bf D}$.

In equation (\ref{VCHE-tnsr}), one recognizes the constitutive
relation for the NS$-\alpha$ model as a variant of the
rate-dependent incompressible homogeneous fluid of second
grade~\cite{DF1974},~\cite{Dunn95}, in which the dissipation,
however, is modified by composition with the Helmholtz operator
$(1-\alpha^2\Delta)$. Thus, the NS$-\alpha$ model has
nearly the same stress tensor as the second-grade fluid. 
However, it is not quite the same. The stress tensor for the
NS$-\alpha$ model uses Navier-Stokes viscous dissipation, instead
of the weaker form of dissipation used for second-grade fluids
that is  independent of wavenumber. Note: despite first
appearances, there is no hyperviscosity in the NS$-\alpha$ model,
only the standard Navier-Stokes viscosity. 

Equations for the second grade fluid
were treated recently in the mathematical
literature~\cite{CV1996},~\cite{CV1997},~\cite{Busuioc}.
Also, in \cite{Shkoller-preprint} local well-posedness results for
some initial value problems for differential type fluids arrising
in non-Newtonian fluids (second-grade and third-grade fluid) were
obtained by transferring the problem from the Eulerian to the
Lagrangian setting, thereby extending the Arnold
\cite{Arnold1966} and Ebin-Marsden program
\cite{Ebin-Marsden1970} to the case of second-grade fluids. For
the case of second-grade fluids, for $k\alpha>1$ the wavenumber
spectrum should still roll over to $k^{-3}$, provided the
Kolmogorov/Oboukov argument still holds for the weaker
dissipation.

The association of turbulence closure models with non-Newtonian
fluids is natural. There is a tradition at least since
Rivlin~\cite{Rivlin57} of modeling turbulence by using continuum
mechanics principles such as objectivity and material frame
indifference (see also~\cite{Chorin88}). For example, this sort
of approach is taken in deriving Reynolds stress algebraic
equation models~\cite{Shih95}. Rate-dependent closure models of
mean turbulence have also been obtained by the two-scale DIA
approach~\cite{Yoshizawa84} and by the renormalization group
method~\cite{Rubinstein90}.

Despite its similarity to the rheology of second-grade fluids, the
alpha parameter in the NS$-\alpha$ model is actually not a
material parameter. Rather, it is a {\bfi flow regime parameter}.
Since the NS$-\alpha$ model describes mean quantities, it was
proposed as a turbulence closure model and this ansatz was
tested by comparing its steady solutions (using the standard
Navier-Stokes viscosity) to mean velocity
measurements in turbulent channel and pipe flows in
\cite{Chen-etal-PRL[1998]}-\cite{Chen-etal-PhysD[1998]}.
These experimental tests show that alpha depends slightly on
Reynolds number and varies slightly with distance from the wall,
for the low to moderate Reynolds numbers available in channel
flow. However, at the high to very high Reynolds numbers
available in pipe flows, alpha becomes independent of Reynolds
number and takes a small value -- about one percent of the pipe
diameter.

%%%%%%%%%%%%%%%%%%%%%%%%%%%%%%%%%%%%%%%%%%%%%%%%%%%%%%%%%%%%%%%%

\section{Guide to recent NS$-\alpha$ model literature} In
\cite{Chen-etal-PRL[1998]}-\cite{Chen-etal-PhysD[1998]} the
authors introduce random fluctuations into the description of the
fluid parcel trajectories in the Lagrangian in Hamilton's
principle for ideal incompressible fluid dynamics. They then take
its statistical average and use the Euler-Poincar\'e theory  of
\cite{HMR-98a}, \cite{HMR-98b}, \cite{HMR-INI-99} to derive
Eulerian closure equations for the corresponding averaged ideal
fluid motions. The Euler-Poincar\'e equation that is used in
deriving the ideal dynamics of the NS$-\alpha$ model is
equivalent in the Eulerian picture to the corresponding
Euler-Lagrange equation for fluid parcel trajectories for
Lagrangians that are invariant under the right-action of the
diffeomorphism group. See \cite{HMR-98a}, \cite{HMR-98b},
\cite{HMR-INI-99}, and references therein for more discussions of
Euler-Poincar\'e equations. The Euler-Poincar\'e theory is
applied for modeling fluctuation effects on 3D Lagrangian mean
and Eulerian mean fluid motion in
\cite{DDH-fluct[1999]}.

The NS$-\alpha$ model equations (also sometimes called in the
literature the viscous Camassa-Holm equations) are proposed in
\cite{Chen-etal-PRL[1998]}-\cite{Chen-etal-PhysD[1998]} as a
turbulence closure approximation for the Navier-Stokes
equations. The analytic form of the velocity profiles based on
the steady NS$-\alpha$ equations away from the viscous sublayer,
but covering at least 95\% of the channel, depends on two free
parameters: the flux Reynolds number $R$, and the wall-stress
Reynolds number $R_0$.  (Due to measurement
limitations, most experimental data are
contained in this region.)  The authors further
reduce the parameter dependence to a single free
parameter by assuming a certain drag law for the wall
friction $D \sim R^2_0/R^2$.  For most of the channel the steady
NS$-\alpha$ solution is shown to be compatible with empirical and
numerical velocity profiles in this subregion.
The NS$-\alpha$ steady velocity profiles agree well outside the
viscous sublayer with data obtained from mean velocity
measurements and simulations of turbulent channel and pipe flow
over a wide range of Reynolds numbers,
\cite{Chen-etal-PRL[1998]}-\cite{Chen-etal-PhysD[1998]}.

%%%%%%%%%%%%%%%%%%%%%%%%%%%%%%%%%%%%%%%%%%%%%%%%%%%%%%%%%%%%%%%%

\paragraph{DNS and comparisons with LES.} In
\cite{Chen-etal[1999]} direct numerical simulations (DNS) of the
NS$-\alpha$ equations are compared with Navier-Stokes DNS and
interpreted as behaving like an LES (Large Eddy Simulation)
model. That is, the  NS$-\alpha$ model is shown to produce an
accurate dynamical description of the large scale features of
turbulence driven at large scales, even at resolutions for which
the fine scales are not resolved.  The NS$-\alpha$-model may at
first appear not to be an LES model, since it has
rate-dependence that is not admitted by LES models. 
However, the NS$-\alpha$ model was recently shown to
transform to a generalized LES similarity model in
\cite{Dom-Holm[2000]}. This is a promising result for LES
modeling, since the mathematical theorems available in \cite{FHTi}
for existence, uniqueness and finite dimensional global attractor
for the NS$-\alpha$ model will now be transferable to the
continuous formulations of this class of generalized LES
similarity models. As a basis for numerical schemes, the
NS$-\alpha$ model also resembles vorticity methods, as observed
in \cite{Shkoller-preprint}, \cite{Leonard},
\cite{Oliver-Shkoller-preprint}. 

%%%%%%%%%%%%%%%%%%%%%%%%%%%%%%%%%%%%%%%%%%%%%%%%%%%%%%%%%%%%%%%%

\paragraph{Geodesic motion.} The completely integrable
one-dimensional Camassa-Holm equation \cite{CH-93} is expressed
on the real line as
\begin{equation}\label{CH-eqn-1D}
\frac{\partial v}{\partial t}
+
u\frac{\partial v}{\partial x}
+
2v\frac{\partial u}{\partial x}
=
0
\,,\quad
u(x,t) = \frac{1}{2} \int_{-\infty}^\infty
e^{|(x-y\,|}\,v(y,t)\,dy
\,.
\end{equation}
Thus, we have $v=u-\partial^2 u/\partial x^2$, cf. equations
(\ref{NS-alpha-eqns}) with definition (\ref{g-inv-alpha}). This
one-dimensional equation is formally the Euler-Poincar\'e equation
for geodesic motion on the diffeomorphism group with respect to
the metric given by the mean kinetic energy Lagrangian, which is
right-invariant under the action of the diffeomorphism group. See
\cite{HMR-98a}, \cite{HMR-98b} for detailed discussions,
applications and references to Euler-Poincar\'e equations of this
type for ideal fluids and plasmas. See \cite{HMR-98a},
\cite{HMR-98b} for the original derivation of the $n$-dimensional
Camassa-Holm, or Euler$-\alpha$ equation in
Euclidean space. See \cite{5-author-99}, \cite{Shkoller-JFA-98}
for discussions of its generalization to Riemannian manifolds,
its existence and uniqueness on a finite time
interval, and more about its relation to the
theory of second grade fluids.  Additional
properties of the Euler$-\alpha$ equations, such as
smoothness of the geodesic spray (the
Ebin-Marsden theorem) are also shown to hold in
\cite{Shkoller-JFA-98} and the limit of zero viscosity for the
corresponding viscous equations is shown to be a
regular limit, even in the presence of boundaries
for either homogeneous (Dirichlet) boundary conditions,
or for boundary conditions involving the
second fundamental form of the boundary. Functional-analytic
studies of the ideal Euler$-\alpha$ model are made in
\cite{Shkoller-preprint}, \cite{MRS-99}. The methods
introduced and applied in \cite{Shkoller-preprint},
\cite{MRS-99}, while geometrical in nature, also address 
analytical issues and obtain results that were not previously
available by traditional techniques for partial differential
equations. For example, these methods produce local in time
existence of $C^\infty$ viscosity independent solutions for the
Euler$-\alpha$ equations in $n$ dimensions (and in particular 3D)
for a fluid container that can be an arbitrary Riemannian
manifold with boundary \cite{MRS-99-footnote}.

%%%%%%%%%%%%%%%%%%%%%%%%%%%%%%%%%%%%%%%%%%%%%%%%%%%%%%%%%%%%%%%%

\paragraph{Mathematical estimates for strong solutions with
dissipation.} Paper \cite{FHTi} provides the mathematical
estimates that are needed to show that the solutions of the
NS-$\alpha$ model exist globally, are unique and possess a global
attractor with finite fractal dimension satisfying equation
(\ref{FractalDimBd}). The estimates in \cite{FHTi} do depend on
the viscosity remaining positive.

\section*{Acknowledgements} We are grateful to S. Chen, J. A.
Domaradzki, L. Margolin, J. E. Marsden, E. Olson, T. Ratiu, S.
Shkoller, J. Tribbia, B. Wingate and S. Wynne for many
constructive comments and enlightening discussions. DDH is also
indebted to G. Eyink, A. Leonard, K. Moffat and D. Thomson for
insightful discussions during our time together (June 1999) at
the Turbulence Programme of the Isaac Newton Institute for
Mathematical Sciences at Cambridge. Research by CF was supported
in part by the National Science Foundation grant DMS-9706903.
Research by DDH was supported by the U.S. Department of Energy
under contracts W-7405-ENG-36 and the Applied Mathematical
Sciences Program KC-07-01-01. The work of EST was supported in
part by the National Science Foundation grants DMS-9704632 and
DMS-9706964.

%%%%%%%%%%%%%%%%%%%%%%%%%%%%%%%%%%%%%%%%%%%%%%%%%%%%%%%%%%%%%%%%%%%


\begin{thebibliography}{99}


\singlespace

\bibitem{MGermano[1992]} M. Germano,
Turbulence: the filtering approach,
{\it J. Fluid Mech.} {\bf238} (1992) 325-336.

\bibitem{SandipGhosal[1999]} S. Ghosal,
Mathematical and physical constraints on large-eddy simulation
of turbulence, {\it AIAA J.} {\bf37} (1999) 425.

\bibitem{Chen-etal-PRL[1998]}
S. Chen, C. Foias, D.D. Holm, E. Olson, E.S. Titi, S. Wynne,
The Camassa-Holm equations as a closure
model for turbulent channel and pipe flow,
{\it Phys. Rev. Lett.} {\bf81} (1998) 5338-5341.

\bibitem{Chen-etal-PF[1998]}
S. Chen, C. Foias, D.D. Holm, E. Olson, E.S. Titi, S. Wynne,
A connection between the Camassa-Holm equations and turbulence
flows in pipes and channels,
{\it Phys. Fluids}, {\bf11} (1999) 2343-2353.

\bibitem{Chen-etal-PhysD[1998]}
S. Chen, C. Foias, D.D. Holm, E. Olson, E.S. Titi, S. Wynne,
The Camassa-Holm equations and turbulence in pipes and
channels, {\it Physica D}, {\bf133} (1999) 49-65.

\bibitem{FHTi}  C. Foias, D. D. Holm and E.S. Titi,
The three dimensional viscous Camassa--Holm equations,
 and their relation to the Navier--Stokes equations and
turbulence theory,
{\it Journal of Dynamics and Differential Equations}, submitted.

\bibitem{Chen-etal[1999]}
S.Y. Chen, D.D. Holm, L.G. Margolin and R. Zhang,
Direct numerical simulations of the Navier-Stokes alpha model,
{\it Physica D}, {\bf133} (1999) 66-83.

\bibitem{Dom-Holm[2000]}
J. A. Domaradzki and D. D. Holm, 
Navier-Stokes-alpha model: 
LES equations  with nonlinear dispersion, 
Special LES volume of {\it ERCOFTAC Bulletin}, to appear (2000).

\bibitem{HMR-98a}  D.D. Holm, J.E. Marsden, T.S. Ratiu,
Euler-Poincar\'e Models of Ideal Fluids with Nonlinear
Dispersion,
{\it Phys. Rev. Lett.} {\bf 80} (1998) 4173-4176.

\bibitem{HMR-98b}   D.D. Holm, J.E. Marsden, and T.S. Ratiu,
Euler-Poincar\'e equations and semidirect
products with applications to continuum theories,
{\it Adv. in Math.} {\bf 137} (1998) 1-81.

\bibitem{Leray[1934]} J. Leray,
Sur le mouvement d'un liquide visqueux emplissant l'espace,
{\it Acta Math.} {\bf63} (1934)
193-248.  Reviewed, e.g., in P. Constantin, C. Foias,  B.
Nicolaenko and R. Temam, {\it Integral manifolds and inertial
manifolds for dissipative partial differential equations}.
Applied Mathematical Sciences, {\bf70}, (Springer-Verlag, New
York-Berlin, 1989).

\bibitem{Gallavotti[1992]} G. Gallavotti, Some rigorous results
about 3D Navier-Stokes, in Les Houches 1992 NATO-ASI meeting on
{\it Turbulence in Extended Systems}, eds. R. Benzi, C. Basdevant
and S. Ciliberto (Nova Science, New York, 1993) pp. 45-81.
(We are grateful to G. Eyink for pointing out this rewference to
us.)

\bibitem{Kuz'min-1983} G. A. Kuz'min,
Ideal incompressible hydrodynamics in terms of the vortex
momentum density, {\it Phys. Lett. A} {\bf96} (1983) 88-90.

\bibitem{Oseledets-1989} V. I. Oseledets,
New form of the Navier-Stokes equation. Hamiltonian formalism,
(in Russian) {\it Moskov. Matemat. Obshch.} {\bf44} no. 3 (267)
(1989) 169-170.

\bibitem{Gama-Frisch[1993]} S. Gama and U. Frisch, Local
helicity, a material invariant for the odd-dimensional
incompressible Euler equations,
in {\it Proceed. NATO-ASI: Theory of Solar and Planetary
Dynamos}, ed.
M. R. E. Proctor, P. C. Mathews and A. M. Rucklidge,
Cambridge University Press (1993), pp.115-119.

\bibitem{HK83} D. D. Holm and B. A. Kupershmidt,
Poisson brackets and Clebsch representations
for magnetohydrodynamics, multifluid  plasmas, and elasticity,
{\it Physica D} {\bf 6} (1983) 347--363.

\bibitem{Foias-Temam[1989]}
C. Foias and R. Temam,
Gevrey class regularity for the
solutions of the Navier-Stokes equations,
{\it J. of Funct. Anal.} {\bf 87} (1989) 359-369.

\bibitem{Doering-Titi[1995]}
C. R. Doering and E. S. Titi,
Exponential  decay rate of the power spectrum for
solutions of the Navier-Stokes equations,
{\it Phys. Fluids} {\bf 7} (1995) 1384-1390.

\bibitem{DDH-fluct[1999]} D.D. Holm,
Fluctuation effects on 3D Lagrangian mean
and Eulerian mean fluid motion,
{\it Physica D}, {\bf133} (1999) 215-269.

\bibitem{Shkoller-preprint} S. Shkoller, The geometry and analysis
of non-Newtonian fluids and vortex methods. Preprint (1999).

\bibitem{Kraichnan[1967]} R. H. Kraichnan,
Inertial ranges in two-dimensional turbulence,
{\it Phys. Fluids} {\bf10} (1967) 1417-1423.

\bibitem{Eyink-Thomson} A version of this scaling
argument wa suggested to one of the authors (DDH) by
G. Eyink and D. Thomson.

\bibitem{Foias97}
C. Foias,
What do the Navier-Stokes equations tell us about turbulence?
in {\it Harmonic analysis and nonlinear differential equations}
(Riverside, CA, 1995),
{\it Contemp. Math.}, {\bf208} (1997) 151--180.

\bibitem{Chen-etal-00}
S.Y. Chen, D.D. Holm, L.G. Margolin and R. Zhang,
Direct numerical simulations of the Navier-Stokes alpha model in
the limit $\alpha\to\infty$. In preparation.

\bibitem{Oboukov} A. M. Oboukov, On the distribution of energy in
the spectrum of a turbulent flow, {\it Dokl. Akad. Sci. Nauk
SSSR} {\bf32A} (1941) 22-24.

\bibitem{Lesieur} M. Lesieur,
{\it	Turbulence in Fluids}, Fluid Mechanics and Its Applications
{\bf40}, Kluwer Academic Publishers: London, 3rd Edition, (1997)
p. 179.

\bibitem{Leith-Kraichnan-70's} C. E. Leith and R. H. Kraichnan,
Predictability of turbulent flows,
{\it J. Atmos. Sci.} {\bf29} (1972) 1041-1058.

\bibitem{DF1974}
J.E. Dunn and R.L. Fosdick,
Thermodynamics, stability, and boundedness
of fluids of complexity 2 and fluids of second grade.
{\it	Arch. Rat. Mech. Anal.} {\bf 56} (1974) 191--252.

\bibitem{Dunn95}
	J.E. Dunn and K.R. Rajagopal,
Fluids of differential type:
    critical reviews and thermodynamic analysis,
{\it	Int. J. Engng. Sci.} {\bf 33} (1995) 689--729.

\bibitem{CV1996}
D. Cioranescu and V. Girault,
Solutions variationelles et classiques d'une famille de fluides
de grade deux.
{\it	C. R. Acad. Sc. Paris} S\'erie 1, {\bf 322} (1996)
1163-1168.

\bibitem{CV1997}
D. Cioranescu and V. Girault,
Weak and classical solutions of a family of second grade fluids.
{\it Int. J. Non-Lin. Mech.} {\bf 32} (1997) 317-335.

\bibitem{Busuioc} V. Busuioc, On second grade fluids with
vanishing viscosity, {\it Compt. Rend. Acad.
Sci. Serie I-Math.} {\bf 328} (1999) 1241-1246.

\bibitem{Arnold1966} V. I. Arnold,
Sur la g\'{e}ometrie differentielle des groupes de Lie de
dimenson infinie et ses applications \`{a} l'hydrodynamique des
fluids parfaits, {\it Ann. Inst. Fourier, Grenoble\/} {\bf 16}
(1966) 319-361.

\bibitem{Ebin-Marsden1970} D. Ebin and J. E. Marsden,
Groups of diffeomorphisms and the motion of an incompressible
fluid, {\it Ann. of Math.} {\bf92} (1970) 102-163.

\bibitem{Rivlin57}
	R.S. Rivlin,
The relation between the flow of non-Newtonian
    fluids and turbulent Newtonian fluids,
{\it	Q. Appl. Math.} {\bf15} (1957) 212-215.

\bibitem{Chorin88}
	A.J. Chorin,
Spectrum, dimension, and polymer analogies in
        fluid turbulence,
{\it		Phys. Rev. Lett.} {\bf 60} (1988) 1947-1949.

\bibitem{Shih95}
	T.H. Shih, J. Zhu, and J.L. Lumley,
A new Reynolds stress algebraic equation model,
{\it		Comput. Methods Appl. Mech. Engrg.} {\bf 125} (1995)
287-302.

\bibitem{Yoshizawa84}
 	A. Yoshizawa,
  Statistical analysis of the derivation of the
  Reynolds stress from its eddy-viscosity representation,
	{\it	Phys. Fluids} {\bf 27} (1984) 1377-1387.

\bibitem{Rubinstein90}
	R. Rubinstein and J.M. Barton,
Nonlinear Reynolds stress models and the
 renormalization group,
	{\it	Phys. Fluids A} {\bf 2} (1990) 1472-1476.

\bibitem{HMR-INI-99}  D.D. Holm, J.E. Marsden, T.S. Ratiu,
The Euler-Poincar\'e Equations in Geophysical Fluid Dynamics,
in {\it Proceedings of the Isaac Newton Institute Programme on
the Mathematics of Atmospheric and Ocean Dynamics}, Cambridge
University Press, to appear. (See section 3).

\bibitem{Leonard} A. Leonard, private communication, June 1999.

\bibitem{Oliver-Shkoller-preprint} M. Oliver and S. Shkoller, The
vortex blob method as a second-grade non-Newtonian fluid.
Preprint (1999).

\bibitem{CH-93}  R. Camassa and D.D. Holm,
An integrable shallow water equation with
peaked solitons,
{\it Phys. Rev. Lett.} {\bf 71} (1993) 1661--1664.

\bibitem{5-author-99} D.D. Holm, S. Kouranbaeva, J.E.
Marsden, T. Ratiu and S. Shkoller,
A nonlinear analysis of the averaged Euler equations.
Unpublished.

\bibitem{Shkoller-JFA-98}  S. Shkoller, Geometry and curvature
of diffeomorphism groups with $H^1$ metric and
mean hydrodynamics,
{\it J. Func. Anal.} {\bf160} (1998) 337-355.

\bibitem{MRS-99}  J.E. Marsden, T. Ratiu and S. Shkoller,
The geometry and analysis of the
averaged Euler equations  and a new
diffeomorphism group,
{\it Geom. Func. Anal.}, to appear.

\bibitem{MRS-99-footnote} In \cite{MRS-99} the Euler$-\alpha$
model is termed the ``averaged Euler equations.''

\end{thebibliography}
\end{document}